\documentclass[conference]{IEEEtran}
\IEEEoverridecommandlockouts

\usepackage{cite}
\usepackage{amsmath,amssymb,amsfonts}
\usepackage{graphicx}
\usepackage{textcomp}
\usepackage{xcolor}
\usepackage{hyperref}
\usepackage{balance}
\usepackage{url}
\usepackage{mathtools, nccmath}
\usepackage{subcaption}
\usepackage{float}
\usepackage{multirow}
\newcommand{\amax}[1]{\text{amax}{#1}}
\usepackage{algorithm}
\usepackage{algpseudocode}
\usepackage{booktabs}
\usepackage{graphicx}
\usepackage{subcaption}
\usepackage{placeins}
\algrenewcommand\algorithmicrequire{\textbf{Require:}}
\algrenewcommand\algorithmicensure{\textbf{Ensure:}}

\def\BibTeX{{\rm B\kern-.05em{\sc i\kern-.025em b}\kern-.08em
    T\kern-.1667em\lower.7ex\hbox{E}\kern-.125emX}}
\begin{document}

\newcommand{\name}{GeoFP8}

\title{\name{}: Geometry-Aware FP8 Gradient Compression for Distributed LLM Training}

\author{
\IEEEauthorblockN{
Jieying Wang\textsuperscript{1},
Zizhong Wang\textsuperscript{3},
Fangru Linghu\textsuperscript{2},
Shuyuan Fan\textsuperscript{2},
Jiajia Li\textsuperscript{3},
and Zhao Zhang\textsuperscript{2}
}

\IEEEauthorblockA{\textsuperscript{1}Department of Computer Science, Rutgers University, Piscataway, NJ, USA}

\IEEEauthorblockA{\textsuperscript{2}Department of Electrical and Computer Engineering, Rutgers University, Piscataway, NJ, USA}

\IEEEauthorblockA{\textsuperscript{3}Department of Computer Science, North Carolina State University, Raleigh, NC, USA}
}

\maketitle

\begin{abstract}
Gradient communication is a primary scaling bottleneck in large language model (LLM) pretraining.
Communicating gradients in low-precision formats, such as FP8 and NVFP4, can significantly reduce the communication volume.
Existing methods quantize gradients via linear or nonlinear mappings in Euclidean space, often degrading model performance because highly anisotropic gradients incur direction-dependent distortion.
We present \name{}, a geometry-informed gradient scaling method that performs low-precision communication in geometry-aware coordinates.
By transforming gradients into a near-isotropic space before quantization, \name{} makes low-precision representations substantially more faithful to their high-precision counterparts.
\name{} only changes the coordinate system used for low-precision gradient communication and does not change the optimizer, training recipe, communication collective, or low-precision format.
We also develop a simplified geometry-aware transformation algorithm with low-rank approximation and selective application to balance the computation overhead and communication reduction.
We examine the empirical convergence of \name{} using Llama-300M and Llama-600M models.
Our results show that \name{} reduces the end-to-end pretraining time of Llama-600M by 7.6\% on 64 NVIDIA GH200 Superchips, while improving the downstream task preservation profile over direct Euclidean FP8 communication under the same optimizer and communication path.
\end{abstract}

\begin{IEEEkeywords}
large language models, distributed pretraining, gradient communication, communication compression, low-precision communication, geometry-aware communication, K-FAC
\end{IEEEkeywords}

\section{Introduction}
\label{sec:intro}

Gradient communication is a major scaling bottleneck in distributed LLM pretraining~\cite{ben2019demystifying}.
In data-parallel, 3D-parallel, or more complex parallel pretraining strategies, the optimizer needs to communicate gradients across graphics processing units (GPUs).
Standard communication libraries, such as NCCL and MPI, implement the ring allreduce algorithm~\cite{patarasuk2009bandwidth}, whose cost depends on the model size and job size (i.e., the number of GPUs). 
Previous work reports that communication accounts for 40\% of the overall pretraining time for GPT-8.3B pretraining across 128 NVIDIA A100 GPUs~\cite{song2023optimusccefficientlargenlp}.

Reducing communication volume via low-precision gradients is an effective way to lower communication overhead, as the bits saved for each gradient representation directly translate into proportional reductions in bandwidth consumption.
Existing low-precision training techniques, such as FP8-LM~\cite{fp8lm} and COAT~\cite{coat}, mainly apply scaling and quantization in Euclidean tensor coordinates, which can still lead to degraded model performance.
They are designed specifically for the AdamW optimizer~\cite{loshchilov2019decoupledweightdecayregularization}.
FP8-LM uses FP8 format for model weights, activations, gradients, and first moment, while the second moment is stored in 16-bit format for division stability.
COAT uses dynamic range strategies (i.e., $S_X=\frac{\amax(X_{\text{FP32}})}{\Delta^{\text{E4M3}}_{\text{MAX}}}$) and mixed-granularity activation quantization to quantize second moment and activation, while the gradients are still in 16-bit format for training stability.
SDP4Bit~\cite{jia2024sdp4bit4bitcommunicationquantization} leverages the sharded data parallel model placement and implements 4-bit gradient communication via Fourier transform and hierarchical all-to-all communication.
However, neither FP8-LM nor COAT achieves comparable model performance to the BF16 baseline. 
FP8-LM shows that seven out of ten downstream tasks perform worse than the baseline.
COAT has only one with a higher score than the baseline among the four downstream tasks.
SDP4Bit only evaluates model performance using validation loss.

\begin{figure}[t]
    \centering
    \includegraphics[width=0.98\linewidth]{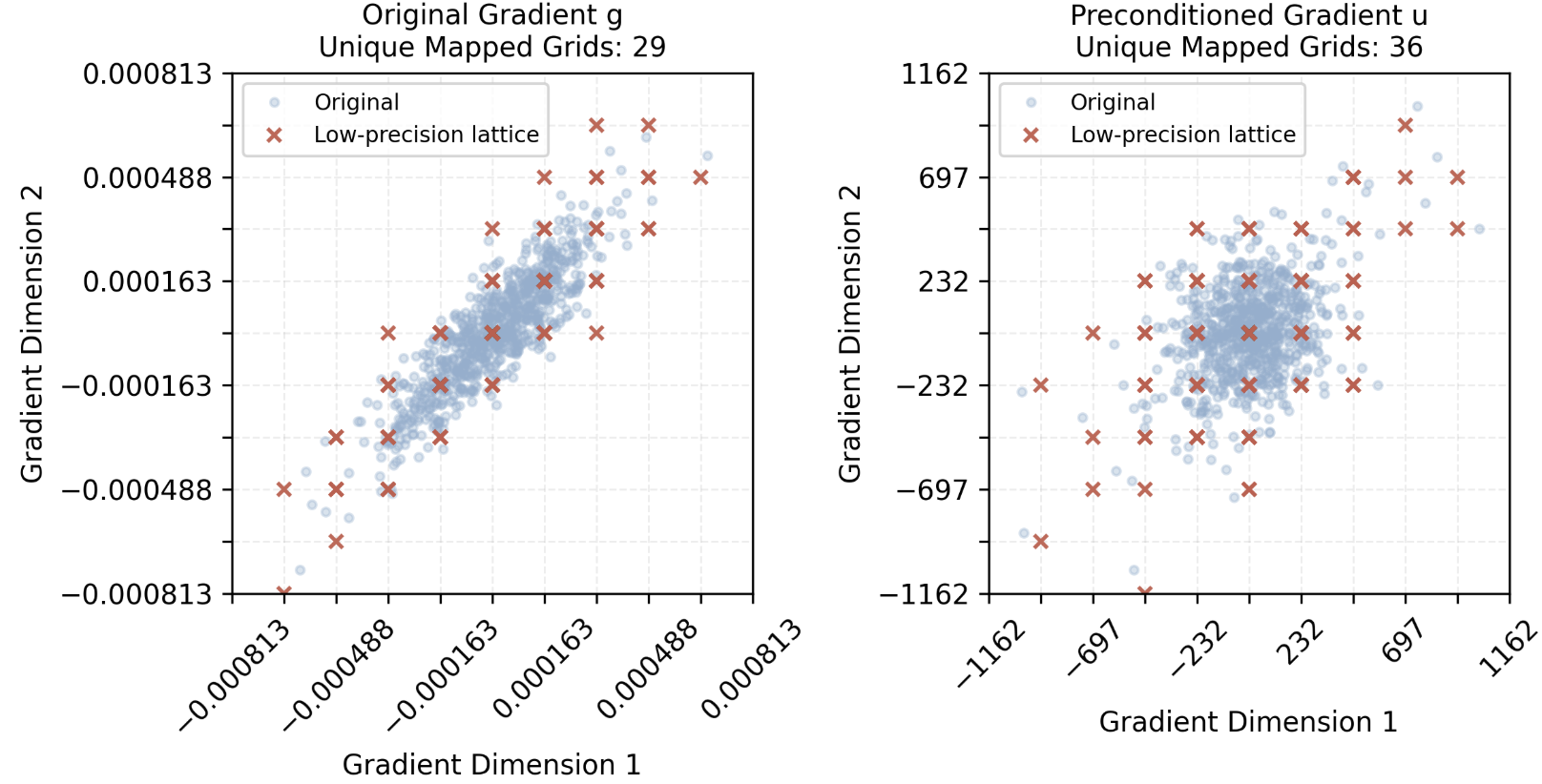}
    \caption{Two-dimensional projection of real gradient samples in different coordinate systems. In the original Euclidean parameter space (left), the gradients are highly elongated and anisotropic. After a geometry-aware transform (right), the same gradients become more rounded and less correlated, making low-precision quantization less sensitive to direction.}
    \label{fig:contour}
\end{figure}

Most existing works propose holistic quantization approaches that quantize not only gradients but also model parameters and optimizer states. 
It is challenging to isolate the impact of gradients in low-precision formats and to justify the additional computational overhead and the reduction in communication.
In this work, we focus exclusively on gradient communication using the FP8 format (native FP8 support was added in NCCL 2.24 in early 2025).
We systematically analyze the optimization process using gradients in FP8 for the Muon optimizer~\cite{jordan2024muon} in Llama model training. 
\name{} only changes the coordinate system used for low-precision gradient communication; it does not change the optimizer, model, training recipe, communication collective, or FP8 format.
We observe that gradients are often highly anisotropic, so direct quantization in Euclidean space can distort some directions more than others, causing the synchronized update to drift from its high-precision counterpart.
As shown in \autoref{fig:contour}, applying a uniform scaling factor on both axes in the Euclidean space causes different error levels.
However, transforming the gradients onto a Riemannian manifold via Fisher whitening yields a distribution that is close to isotropic.

Inspired by this observation, we propose \name{}, a practical geometry-informed gradient scaling technique that improves the fidelity--efficiency tradeoff of FP8 gradient communication.
The key idea is simple: quantize gradients in geometry-aware coordinates rather than raw Euclidean coordinates, so scaling to FP8 introduces comparable errors across dimensions.
\name{} is enabled by practical local geometry approximations based on K-FAC~\cite{amari_iga_book,amari1998natural,martens2015kfac} (see \S\ref{subsec:background_kfac}), and we refer to the full geometry approximation as K-FAC in the following discussion.
We use this approximation only to construct temporary communication coordinates, not to replace the optimizer update with a K-FAC or natural-gradient step.
Naively applying K-FAC to all gradients yields a better model performance than existing FP8 scaling techniques.
However, the computational complexity of K-FAC makes it less practical, as the additional computation time outweighs the reduced communication time.
So we select the layers most sensitive to quantization and apply simplified K-FAC only to these layers, while other layers use standard Euclidean scaling. 
This selective strategy reduces the computational overhead of K-FAC while preserving the resulting model performance.

We implement \name{} in the Megatron-LM~\cite{megatronlm} framework and pretrain Llama-300M and Llama-600M with the OpenWebText~\cite{openwebtext} dataset.
We compare the pretrained model performance using \name{}, FP32, and a Euclidean baseline (i.e., a layer-wise scaling).
Our error analysis shows that \name{} reduces the error, i.e., the difference between averaged gradients in FP8 and FP32, by up to \textbf{67.4\%} over the Euclidean baseline.
The pretraining studies show that Llama models pretrained with \name{} achieve a better downstream task preservation profile than the Euclidean scaling approach under the same optimizer and communication path.
Compared to gradients in FP32, \name{} reduces the gradient communication volume by \textbf{75.0\%}, resulting in a \textbf{7.6\%} reduction in overall pretraining time across 64 GH200 superchips.
The 75.0\% reduction refers to gradient payload size, while the end-to-end speedup is smaller because total pretraining time also includes forward and backward computation, optimizer updates, quantization and dequantization, geometry transforms, synchronization overheads, and non-gradient communication.
At the same scale, direct Euclidean FP8 achieves a 10.79\% reduction, showing that \name{} trades part of the raw FP8 speedup for improved downstream preservation.

This paper makes three main contributions:
\begin{itemize}
    \item \textbf{A new diagnosis for low-precision scaling error.} We show that the error of Euclidean gradient scaling is a mismatch between anisotropic local gradient geometry and axis-aligned quantization.
    \item \textbf{A selective geometry-aware scaling design.} We develop \textbf{\name{}}, which keeps the standard Euclidean scaling for most layers and applies geometry-aware scaling only where the model performance justifies the cost.
    \item \textbf{A substantially better fidelity--efficiency tradeoff.} We show that \name{} improves the downstream preservation profile over direct Euclidean FP8 while retaining a fraction of the FP8 speedup over FP32, making geometry-aware low-precision communication practical in our Megatron-LM pretraining experiments.
\end{itemize}

The 7.6\% overall time reduction may be considered marginal.
However, modern LLMs are pretrained with tens of thousands of GPUs over several months. 
Even a 7.6\% reduction in time can translate to millions of dollars.
On the other hand, the interconnect in the GPU cluster is a shared resource.
A reduced bandwidth consumption will enable more efficient network sharing.
The communication bottleneck is becoming increasingly prominent as interconnect bandwidth increases more slowly than the compute elements, i.e., the GPUs.
Although our experiments are on FP8, the geometry-transform overhead is mostly independent of the low-precision communication format, so the same principle can be applied to future FP4 communication when software and hardware support mature.
This trend makes geometry-aware communication a promising direction for future lower-bit LLM pretraining systems.

\section{Background}
\label{sec:back}

This section presents the communication setting considered in this paper and the K-FAC-based local metric that later defines the communication coordinates of \name{}. We first summarize how communication arises in data-parallel and 3D-parallel LLM pretraining. We then introduce the layerwise block-diagonal K-FAC approximation used to obtain tractable local geometry~\cite{megatronlm2019,megatronlm,gpipe,pipedream2bw,zero,zeropp,amari1998natural,amari_iga_book,martens2015kfac,kfacmodern}.

\subsection{Communication in Data-Parallel and 3D-Parallel Pretraining}
\label{subsec:background_comm}

Distributed LLM pretraining requires communication whenever model states, activations, or gradients are partitioned across devices. The exact communication pattern depends on the parallelization strategy, but in all cases communication is on the critical path of each pretraining iteration and can become a major scaling bottleneck as model size and system scale grow~\cite{ben2019demystifying,megatronlm2019,megatronlm,song2023optimusccefficientlargenlp}.

\paragraph{Data parallelism}
In data-parallel (DP) pretraining, each worker stores a full replica of the model and processes a different shard of the mini-batch. After the backward pass, the local gradients computed on different workers must be aggregated before the optimizer update so that all replicas remain synchronized~\cite{megatronlm2019,megatronlm,zero}. If worker $r$ produces local gradient $g^{(r)}$, the synchronized gradient is
\begin{equation}
\bar{g}
=
\frac{1}{R}
\sum_{r=1}^{R} g^{(r)},
\label{eq:dp_grad_avg}
\end{equation}
where $R$ is the number of DP workers. In practice, this aggregation is typically implemented by all-reduce, or equivalently by reduce-scatter followed by all-gather in sharded settings such as ZeRO-style pretraining~\cite{patarasuk2009bandwidth,zero,zeropp}. Because the full gradient set is communicated at every iteration, the communication cost grows with both model size and number of workers, making DP gradient communication a primary target for compression~\cite{qsgd,dgc,powersgd,onebitadam,zeropp}.

\paragraph{3D parallelism}
Modern large-scale LLM pretraining often combines data parallelism (DP), tensor parallelism (TP), and pipeline parallelism (PP), which is commonly referred to as 3D parallelism~\cite{megatronlm,megatronlm2019,song2023optimusccefficientlargenlp}. In this setting, communication arises along three distinct paths.

First, in tensor parallelism, individual layers are partitioned across devices, so partial tensor results must be exchanged inside a layer during the forward and backward passes. Depending on the operator and partition rule, this communication is implemented through collectives such as all-reduce, all-gather, or reduce-scatter to assemble outputs or accumulate partial gradients across TP ranks~\cite{megatronlm2019,megatronlm}. Second, in pipeline parallelism, the model is divided into stage-wise partitions, so adjacent stages communicate boundary activations in the forward pass and boundary gradients in the backward pass through point-to-point sends and receives~\cite{gpipe,pipedream2bw,megatronlm}. Third, across DP groups, replicated parameters or optimizer shards still require communication so that the distributed optimizer applies a consistent update~\cite{megatronlm,zero,zeropp}.

Therefore, 3D-parallel pretraining contains multiple communication paths with different semantics: TP communicates intra-layer partial tensor results, PP communicates inter-stage activations and gradients, and DP communicates gradient or optimizer-update information across replicas~\cite{megatronlm,gpipe,pipedream2bw,song2023optimusccefficientlargenlp}. In this paper, we focus on the DP gradient communication path. This choice isolates the communication bottleneck that is most directly compatible with low-precision compression and allows us to study the effect of gradient-coordinate choice while keeping the optimizer semantics unchanged outside the communication step~\cite{qsgd,powersgd,fp8formats,jia2024sdp4bit4bitcommunicationquantization}.

\subsection{K-FAC Block-Diagonal Approximation}
\label{subsec:background_kfac}

The Fisher information matrix provides a local metric for measuring parameter perturbations in neural-network optimization~\cite{amari1998natural,amari_iga_book}. Let $\theta$ denote the model parameters and let $F(\theta)$ denote the Fisher matrix. A perturbation $\Delta\theta$ has local squared length
\begin{equation}
\|\Delta \theta\|_{F}^{2}
=
\Delta \theta^{\top} F(\theta)\, \Delta \theta.
\label{eq:fisher_metric}
\end{equation}
However, directly forming or inverting the full Fisher matrix is infeasible for modern LLMs because it couples all parameters across all layers.

K-FAC makes this local geometry tractable through a structured approximation to the Fisher. In the standard K-FAC formulation, the Fisher is treated through layerwise blocks, and each such block is approximated by a Kronecker product of two smaller factors~\cite{martens2015kfac,kfacmodern}. For a model with weight matrices $\{W_l\}_{l=1}^{L}$, this layerwise approximation can be written as
\begin{equation}
F(\theta)
\approx
\operatorname{blockdiag}
\left(
F_{W_1},
F_{W_2},
\ldots,
F_{W_L}
\right),
\label{eq:kfac_blockdiag}
\end{equation}
where each diagonal block describes the local geometry of one layer. This block-diagonal approximation removes cross-layer curvature terms and makes the metric estimable and applicable on a per-layer basis.

For a linear layer with weight matrix $W \in \mathbb{R}^{d_{\mathrm{out}} \times d_{\mathrm{in}}}$, let $a$ denote the layer input activation and let $\delta$ denote the gradient with respect to the layer pre-activation output. Using the standard column-major vectorization convention, K-FAC approximates the layerwise Fisher block as
\begin{equation}
F_W
\approx
A \otimes G,
\qquad
A = \mathbb{E}[a a^{\top}],
\qquad
G = \mathbb{E}[\delta \delta^{\top}],
\label{eq:kfac_fisher_block}
\end{equation}
where $A \in \mathbb{R}^{d_{\mathrm{in}} \times d_{\mathrm{in}}}$ captures input-side second-order statistics and $G \in \mathbb{R}^{d_{\mathrm{out}} \times d_{\mathrm{out}}}$ captures output-gradient-side second-order statistics~\cite{martens2015kfac,kfacmodern}. Thus, instead of maintaining a full Fisher matrix over all model parameters, K-FAC maintains small structured factors for each layer.

This paper does not use K-FAC as an optimizer and does not apply the natural-gradient update. We use the K-FAC block only as a tractable local metric for defining communication coordinates. The optimizer, model architecture, and distributed training structure remain unchanged; only the coordinate system used during low-precision gradient communication is modified.

\section{Related Work}
\label{sec:related}
This section summarizes related works in three categories:
communication compression, low-precision LLM pretraining, and geometry-aware optimization.

\subsection{Gradient Communication Compression}
A long line of work reduces distributed-training overhead by reducing the communication volume.
Example methods include quantization, sparsification, low-rank approximation, and optimizer-specific communication schemes~\cite{qsgd,dgc,powersgd,onebitadam,zeropp,loco,song2023optimusccefficientlargenlp,jia2024sdp4bit4bitcommunicationquantization}. 
These methods show that gradient communication can often be reduced substantially without fully sacrificing optimization quality, making communication compression a practical systems direction for large-scale pretraining. 
Recent work has further pushed this trend toward lower-bit communication in modern LLM settings, including sharded and hierarchical communication designs~\cite{zeropp,loco,jia2024sdp4bit4bitcommunicationquantization}. 
Our work aligns with existing approaches but differs in its mechanism: rather than modifying the communication pattern, reducing rank, or introducing sparsity, we study how the \emph{coordinate system} used for gradient communication affects the fidelity of low-precision communication.
These methods are important related systems, but they are not the direct controlled baseline for our study. QSGD changes the quantization rule, DGC introduces sparsity, PowerSGD communicates a low-rank gradient representation, and 1-bit Adam is coupled with Adam-style optimizer dynamics. LoCo is closely related because it targets low-bit gradient communication, but its main mechanism is error compensation in the original gradient coordinates. In contrast, \name{} keeps the optimizer, communication collective, FP8 format, and training recipe fixed, and changes only the communication coordinate system.

\subsection{Low-precision LLM Pretraining}
Another closely related line of work studies FP8 and other low-precision formats for efficient pretraining~\cite{mixedprecisiontraining,fp8formats,fp8graphcore,fp8lm,coat}. 
These methods typically rely on scaling a tensor before quantization so that its values better match the representable range of the target format~\cite{fp8formats,fp8lm}. 
In this setting, the main design questions concern how to choose the scaling statistics and how to maintain stable pretraining under reduced precision~\cite{fp8formats,fp8lm,coat}. 
Representative systems such as FP8-LM and COAT demonstrate that low-precision pretraining can be extended to broader parts of the pretraining stack, including gradients, activations, and optimizer states, under carefully designed scaling rules~\cite{fp8lm,coat}. 
SDP4Bit further studies low-bit communication in sharded data-parallel LLM pretraining and shows that communication-aware smoothing and hierarchical quantization can substantially narrow the quality gap relative to higher-precision baselines~\cite{jia2024sdp4bit4bitcommunicationquantization}. 
Its setting changes the sharded communication path and jointly designs the communication algorithm, while our experiments isolate the coordinate choice under the same allreduce path.
Unlike these holistic systems, \name{} focuses only on gradient communication and does not quantize model parameters, activations, or optimizer states. These works establish the practical importance of scaling for low-precision pretraining and communication, but they still operate on tensors in their original Euclidean coordinates.

\subsection{Geometry-aware Optimization and \name{} Position}
Our work is also related to geometry-aware optimization, especially natural gradient and K-FAC-style approximations~\cite{amari1998natural,martens2015kfac,kfacmodern}. 
These methods use local second-order structure to precondition gradients, primarily to improve optimization efficiency or curvature adaptation. 
This is different from using the same structure as a temporary representation for communication.
In contrast, we use the same geometric intuition for a different purpose: improving the \emph{fidelity of low-precision communication}. 
\name{} does not change the optimizer into a second-order method; instead, it uses local geometry to transform gradients into communication coordinates that are more amenable to FP8 quantization. 
The transformed gradient is mapped back before the optimizer update, so the optimizer receives a gradient in the original parameter coordinates.
In this sense, \name{} is complementary to both prior communication-compression methods and geometry-aware optimization methods. 
\name{} differs from existing low-precision techniques in that it explores an alternative coordinate system to apply quantization instead of choosing a better scale in Euclidean space.

\FloatBarrier
\section{Design}
\label{sec:design}

This section presents the design of geometry-informed low-precision gradient communication. The central idea is to keep the standard distributed pretraining and optimizer pipeline unchanged, but to change the coordinate system in which gradient quantization and communication are performed. Instead of communicating weight gradients directly in Euclidean coordinates, \name{} maps them into coordinates induced by the K-FAC local metric, performs low-precision communication in that space, and then maps the synchronized result back before the optimizer update.

\subsection{Geometry-Aware Communication Coordinates}
\label{subsec:design_geometry_coordinates}

We focus on the weight gradient
\begin{equation}
W_g \in \mathbb{R}^{d_{\mathrm{out}} \times d_{\mathrm{in}}}
\label{eq:design_Wg}
\end{equation}
of a linear layer. In conventional low-precision communication, $W_g$ is scaled, quantized, communicated, dequantized, and averaged directly in Euclidean parameter coordinates. This is efficient, but it treats all directions uniformly. When the local gradient geometry is anisotropic, as shown in \autoref{fig:contour}, the same low-precision scale must cover directions with very different magnitudes. As a result, quantization error can become strongly direction-dependent: large-variance directions dominate the scale, while small-variance directions suffer relatively larger distortion.

\paragraph{Local metric from K-FAC}
Following the K-FAC block introduced in Section~\ref{subsec:background_kfac}, the local metric of a linear layer is approximated by
\begin{equation}
F_W
\approx
A \otimes G,
\qquad
A = \mathbb{E}[a a^\top],
\qquad
G = \mathbb{E}[\delta \delta^\top],
\label{eq:design_kfac_metric}
\end{equation}
where $A$ captures input-side second-order statistics and $G$ captures output-gradient-side second-order statistics. For a small perturbation $\Delta W$ of the layer weight, the corresponding K-FAC local squared length is
\begin{equation}
\|\Delta W\|_{F_W}^{2}
=
\operatorname{vec}(\Delta W)^\top
(A \otimes G)
\operatorname{vec}(\Delta W).
\label{eq:design_kfac_norm_vec}
\end{equation}
Using standard Kronecker and trace identities, this can be equivalently written as
\begin{equation}
\|\Delta W\|_{F_W}^{2}
=
\operatorname{tr}
\left(
A \Delta W^\top G \Delta W
\right).
\label{eq:design_kfac_norm_trace}
\end{equation}

Let
\begin{equation}
A = L_A L_A^\top,
\qquad
G = L_G L_G^\top
\label{eq:design_chol}
\end{equation}
be Cholesky factorizations of the regularized K-FAC factors. Substituting these factorizations into Eq.~\eqref{eq:design_kfac_norm_trace} gives
\begin{align}
\|\Delta W\|_{F_W}^{2}
&=
\operatorname{tr}
\left(
L_A L_A^\top
\Delta W^\top
L_G L_G^\top
\Delta W
\right)
\nonumber \\
&=
\operatorname{tr}
\left(
L_A^\top
\Delta W^\top
L_G
L_G^\top
\Delta W
L_A
\right)
\nonumber \\
&=
\left\|
L_G^\top \Delta W L_A
\right\|_F^2 .
\label{eq:design_sphere_proof}
\end{align}
Therefore, the K-FAC ellipsoid
\begin{equation}
\left\{
\Delta W:
\|\Delta W\|_{F_W}^{2}
\leq
\epsilon
\right\}
\label{eq:design_fisher_ellipsoid}
\end{equation}
is mapped by the factor transform
\begin{equation}
U
=
L_G^\top \Delta W L_A
\label{eq:design_metric_coordinate}
\end{equation}
to the Euclidean Frobenius ball
\begin{equation}
\left\{
U:
\|U\|_F^2
\leq
\epsilon
\right\}.
\label{eq:design_euclidean_ball}
\end{equation}
This shows that the K-FAC factors turn the local anisotropic metric into a coordinate system whose metric ball is spherical. Equivalently, directions that are stretched or compressed by the local Fisher geometry become more uniformly scaled after the coordinate transformation.

\paragraph{Whitening the communicated gradient}
The metric-ball argument above explains the geometric role of the K-FAC factors. For communication, we apply the corresponding inverse factors to the gradient before quantization. Specifically, the full geometry-informed communication coordinate is defined as
\begin{equation}
\widetilde{W}_g
=
L_G^{-1} W_g L_A^{-\top}.
\label{eq:design_full_transform}
\end{equation}
This transform can also be viewed as a whitening operation on the layerwise gradient statistics. Under the Fisher/K-FAC approximation, the dominant second-order structure associated with $\operatorname{vec}(W_g)$ is represented by the same Kronecker-structured factors. Therefore,
\begin{equation}
\operatorname{vec}(\widetilde{W}_g)
=
\left(
L_A^{-1}
\otimes
L_G^{-1}
\right)
\operatorname{vec}(W_g)
\label{eq:design_vec_whiten}
\end{equation}
removes the dominant input-side and output-side scaling captured by $A$ and $G$. In the transformed coordinates, the gradient distribution is therefore closer to isotropic, so an axis-aligned low-precision quantization lattice interacts with the gradient more uniformly than it does in the original Euclidean coordinates.

After transformation, \name{} applies the same low-precision communication procedure used by the Euclidean baseline:
\begin{equation}
\widehat{\widetilde{W}}_g
=
\operatorname{Comm}_{\mathrm{FP8}}
\left(
\widetilde{W}_g
\right),
\label{eq:design_fp8_comm}
\end{equation}
where $\operatorname{Comm}_{\mathrm{FP8}}(\cdot)$ denotes scaling, FP8 quantization, collective communication, dequantization, and averaging. The synchronized transformed gradient is then mapped back to Euclidean coordinates:
\begin{equation}
\widehat{W}_g
=
L_G
\widehat{\widetilde{W}}_g
L_A^\top.
\label{eq:design_full_inverse_transform}
\end{equation}
The mapped-back gradient $\widehat{W}_g$ is passed to the same optimizer used by the high-precision and Euclidean low-precision baselines.

Importantly, this procedure changes only the communication coordinates. It does not change the model architecture, the distributed parallelization strategy, or the optimizer update rule. Thus, \name{} is not a K-FAC optimizer or a natural-gradient method; it uses K-FAC-style local geometry only to improve the fidelity of low-precision gradient communication.

\paragraph{Input-side practical form}
The full two-sided transform in Eq.~\eqref{eq:design_full_transform} makes the geometric principle explicit, but applying it to every layer can introduce substantial matrix-computation and memory overhead in large-scale LLM pretraining. In practice, \name{} uses an input-side version of the same idea:
\begin{equation}
\widetilde{W}_g
=
W_g L_A^{-\top},
\label{eq:design_input_side_transform}
\end{equation}
followed by FP8 communication in the transformed coordinates and inverse mapping
\begin{equation}
\widehat{W}_g
=
\widehat{\widetilde{W}}_g L_A^\top.
\label{eq:design_input_side_inverse}
\end{equation}
This input-side form preserves the most useful part of the local geometry while avoiding the cost of maintaining and applying both K-FAC factors. It can also be interpreted as turning the input-side component of the K-FAC metric into a more isotropic coordinate system. Specifically, if only the input-side metric is used, then
\begin{equation}
\operatorname{tr}
\left(
A \Delta W^\top \Delta W
\right)
=
\left\|
\Delta W L_A
\right\|_F^2,
\label{eq:design_input_side_sphere}
\end{equation}
so the input-side ellipsoidal geometry is again mapped to a Euclidean Frobenius ball.

The full two-sided formulation in Eq.~\eqref{eq:design_full_transform} exposes the geometric principle, while the input-side form in Eq.~\eqref{eq:design_input_side_transform} gives the practical communication coordinate used by \name{}. The remaining system questions are how much of this geometry is needed in practice, how accurately the input-side factor must be represented, and where the geometry-aware path should be applied. Section~\ref{sec:implementation} answers these questions empirically and turns the coordinate transform into the final system through three simplifications: retaining only the input-side geometry, approximating it with a low-rank factor, and applying it selectively to the most numerically vulnerable layers.

\FloatBarrier
\section{Implementation}
\label{sec:implementation}

\begin{figure}[t]
    \centering
    \includegraphics[width=\linewidth]{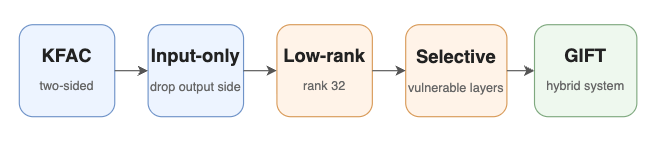}
    \caption{Roadmap of the practical simplifications used to turn the full geometry-informed communication formulation into the final GeoFP8 system.}
    \label{fig:design_roadmap}
\end{figure}

This section turns the full geometry-informed communication formulation of Section~\ref{sec:design} into a practical system implementation.
Guided by the roadmap in Figure~\ref{fig:design_roadmap}, our implementation answers four empirical questions:
Which part of the full transform matters most?
How accurately must the remaining geometry be represented?
Which layers benefit sufficiently to justify the extra cost?
And how should the resulting design be integrated into a standard distributed pretraining pipeline?

The final system answers these questions through three design decisions:
(\emph{i}) retain only the input-side geometry,
(\emph{ii}) approximate it with a rank-32 low-rank representation, and
(\emph{iii}) deploy it only on the most numerically vulnerable layers.
These decisions are made sequentially: each step removes a source of overhead from the full geometry-informed formulation only after the reconstruction study shows that the corresponding fidelity loss is small.

\subsection{Practical Simplification I: Retaining Only the Input-Side Geometry}
\label{subsec:impl_simplifications}

The full two-sided formulation in Section~\ref{sec:design} is a useful conceptual starting point, but applying both sides of the transform to every eligible layer is too expensive for large-scale end-to-end pretraining.
We therefore begin by asking which side of the transform is actually responsible for the practical fidelity gain.

To answer this question, we perform a controlled one-step FP8 round-trip analysis in Euclidean gradient space.
For each method, we start from an FP32 gradient, map it into the corresponding communication coordinates, apply an FP32$\rightarrow$FP8$\rightarrow$FP32 round trip there, map the result back to Euclidean gradient space, and compare it against the original FP32 gradient.

We report four reconstruction metrics:
\begin{equation}
\begin{aligned}
\mathrm{RelL2} &= \frac{\|\hat{g}-g\|_2}{\|g\|_2},\\
\mathrm{Cos} &= \frac{\langle \hat{g}, g\rangle}{\|\hat{g}\|_2\,\|g\|_2},\\
\mathrm{MaxErr} &= \|\hat{g}-g\|_\infty,\\
\mathrm{MSE} &= \frac{\|\hat{g}-g\|_2^2}{n},
\end{aligned}
\label{eq:recon_metrics}
\end{equation}
where $g$ is the original FP32 gradient, $\hat{g}$ is the reconstructed gradient after the FP8 round trip, and $n$ is the number of entries in the gradient tensor.
Cos measures whether the FP8 round trip preserves the update direction, while RelL2, MaxErr, and MSE measure the magnitude and worst-case size of the communication perturbation.
These metrics therefore provide a mechanism-level fidelity test before running full end-to-end pretraining.
Table~\ref{tab:full_vs_one_sided} reports the results.

\begin{table}[!t]
\centering
\caption{One-step FP8 round-trip reconstruction error for the full and one-sided transforms, measured in Euclidean gradient space against the original FP32 gradient. Lower is better for RelL2, MaxErr, and MSE; higher is better for Cos.}
\label{tab:full_vs_one_sided}
\resizebox{\linewidth}{!}{
\begin{tabular}{lcccc}
\toprule
Method & RelL2 & Cos & MaxErr & MSE \\
\midrule
Euclidean baseline    & 5.281973e-02 & 9.986041e-01 & 1.625657e-04 & 1.289936e-10 \\
Output-side only          & 5.282879e-02 & 9.986036e-01 & 1.629426e-04 & 1.290378e-10 \\
Input-side only           & 1.768108e-02 & 9.998500e-01 & 5.332805e-05 & 1.445418e-11 \\
K-FAC                      & 1.770532e-02 & 9.998502e-01 & 5.332805e-05 & 1.449384e-11 \\
\bottomrule
\end{tabular}
}
\end{table}

The result is clear: the output-side transform provides only a marginal contribution over the Euclidean baseline, whereas the input-side transform already matches the full two-sided transform almost exactly across all metrics.
This shows that, in our profiled setting, nearly all of the measured FP8 round-trip fidelity benefit of full geometry-aware communication is captured by the input-side geometry alone.

We therefore remove the output-side transform entirely and retain only the input-side term.
The resulting communication coordinates can be written as
\begin{equation}
\widetilde{W}_g = W_g T_A,
\label{eq:impl_right_only_transform}
\end{equation}
where $T_A$ is an input-side transform derived from the factor $A$.
FP8 quantization and communication are then performed on $\widetilde{W}_g$, and the synchronized result is mapped back to Euclidean gradient space by the corresponding inverse right-side transform.
This simplification is the first key step that makes geometry-informed communication practical.

\subsection{Practical Simplification II: Low-Rank Approximation of the Input-Side Geometry}
\label{subsec:impl_lowrank}

Once the full transform is reduced to the input side, the next question is how accurately the remaining geometry must be represented.
Using the full factor $A$ for every selected layer is still costly, so we seek a lower-cost approximation that preserves most of its communication benefit.

Let
\begin{equation}
A \approx \mathbb{E}[x x^\top], \qquad A \succeq 0,
\label{eq:impl_A_psd}
\end{equation}
and let
\begin{equation}
A \approx U_r \Lambda_r U_r^\top
\label{eq:impl_A_lowrank}
\end{equation}
be a rank-$r$ approximation with $r \ll d_{\text{in}}$.
We construct the input-side communication transform from this reduced representation and again evaluate the resulting methods through the same Euclidean-space FP8 round-trip analysis.
Table~\ref{tab:input_side_ablation} reports the comparison.

\begin{table}[!t]
\centering
\caption{Ablation of input-side approximations, evaluated in Euclidean gradient space after FP8 round-trip. Low-rank approximations recover nearly all of the benefit of the full input-side transform, and rank~32 provides the best practical cost--quality tradeoff.}
\label{tab:input_side_ablation}
\resizebox{\linewidth}{!}{
\begin{tabular}{lcccc}
\toprule
Method & RelL2 & Cos & MaxErr & MSE \\
\midrule
Euclidean baseline      & 5.281973e-02 & 9.986041e-01 & 1.625657e-04 & 1.289936e-10 \\
Input-side full-$A$         & 1.768108e-02 & 9.998500e-01 & 5.332805e-05 & 1.445418e-11 \\
Input-side diag-$A$         & 5.262250e-02 & 9.986144e-01 & 1.609139e-04 & 1.280321e-10 \\
Input-side block-$A$-32     & 3.168243e-02 & 9.994993e-01 & 1.269338e-04 & 4.641014e-11 \\
Input-side block-$A$-64     & 2.754295e-02 & 9.996220e-01 & 1.051468e-04 & 3.507491e-11 \\
Input-side block-$A$-128    & 2.431718e-02 & 9.997058e-01 & 1.051468e-04 & 2.734022e-11 \\
Input-side low-rank-$A$-8   & 1.958074e-02 & 9.998106e-01 & 7.036887e-05 & 1.772696e-11 \\
Input-side low-rank-$A$-16  & 1.731027e-02 & 9.998523e-01 & 5.883700e-05 & 1.385426e-11 \\
Input-side low-rank-$A$-32  & 1.720702e-02 & 9.998551e-01 & 6.311573e-05 & 1.368948e-11 \\
\bottomrule
\end{tabular}
}
\end{table}

Two conclusions follow from Table~\ref{tab:input_side_ablation}.
First, a diagonal approximation barely improves over the Euclidean baseline, indicating that coarse per-dimension scaling is insufficient to capture the useful geometry.
Second, low-rank approximations recover nearly all of the benefit of the full input-side transform.
Among them, rank~32 is already nearly identical to the full-$A$ result while being substantially cheaper to store and apply.
We use rank~32 as the operating point for our implementation rather than as a model-independent constant.
We therefore adopt \textbf{input-side only, low-rank-32} as the geometry-informed transform in the final implementation.

\subsection{Practical Simplification III: Selective Deployment on Vulnerable Layers}
\label{subsec:impl_selective}

Even after the first two simplifications, applying geometry-informed communication to every layer would still introduce unnecessary overhead.
The next question is, therefore, whether all layers need the geometry-aware branch uniformly.

To answer this question, we profile the Euclidean baseline during a 32-GPU pretraining run of a 600M-parameter Llama-style model and measure the numerical vulnerability of each MLP layer (both fc1 and fc2) over the first 100 training steps.
For each profiled module, we flatten its direct parameter gradients, compute a synchronized FP8 scale, encode the flattened gradient into FP8, and count how often encoded values hit the upper and lower FP8 boundaries.
These boundary-hit ratios are accumulated separately for fc1 and fc2 layers and then averaged over the profiling window.
We rank layers by the sum of their average upper- and lower-boundary hit ratios, which serves as our numerical-vulnerability score.

\begin{figure}[t]
    \centering
    \includegraphics[width=0.92\linewidth]{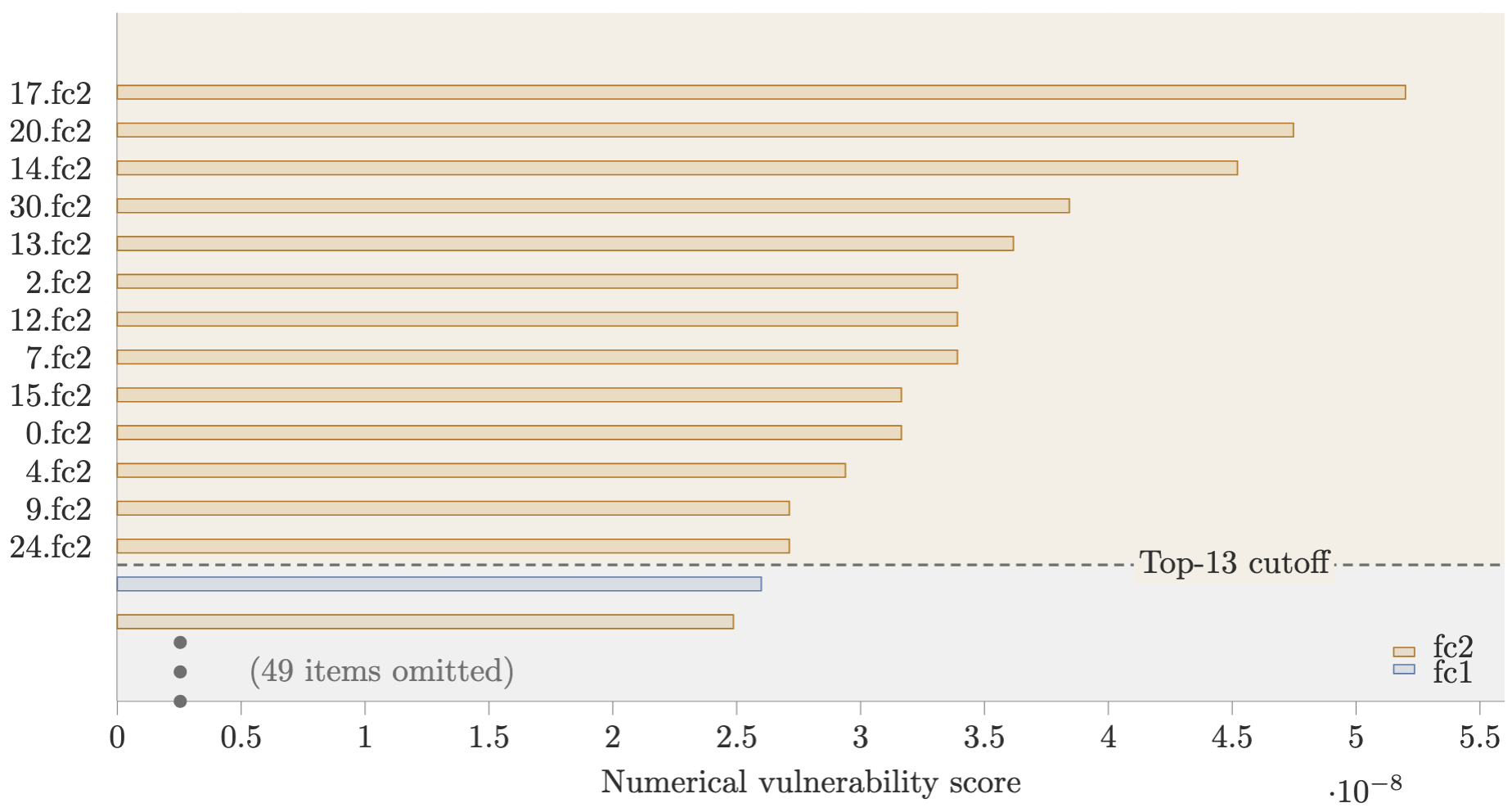}
    \caption{Ranking MLP layers by average numerical vulnerability under the Euclidean baseline during the first 100 training steps. In this profile, the top 13 vulnerable layers are all fc2 layers, and the visible drop after this group motivates selective deployment of the geometry-informed branch for this model and recipe.}
    \label{fig:layer_overflow_rank}
\end{figure}

Figure~\ref{fig:layer_overflow_rank} shows the resulting profile.
The vulnerability is highly concentrated: after the top 13 entries, the score drops visibly, suggesting that the most problematic FP8 distortion is confined to a relatively small subset of layers.
In this profile, the top 13 vulnerable layers are all fc2 layers, while the first fc1 layer appears immediately after this group.
Thus, the number 13 is not intended as a model-independent constant; it is the operating point selected by the vulnerability-ranking procedure for this model and recipe.
This motivates a simple selective-deployment policy:
we enable the geometry-informed branch only for these selected vulnerable MLP layers and keep the Euclidean baseline for all remaining layers.
This policy concentrates additional computation where geometry-aware communication is most likely to help, while preserving the fast path for most of the model.
For a new model architecture or training recipe, the same profiling procedure can be rerun automatically to select the layer set.

\subsection{Pretraining-Time Pipeline}
\label{subsec:impl_pipeline}
\begin{figure}[!t]
    \centering
    \includegraphics[width=0.94\linewidth]{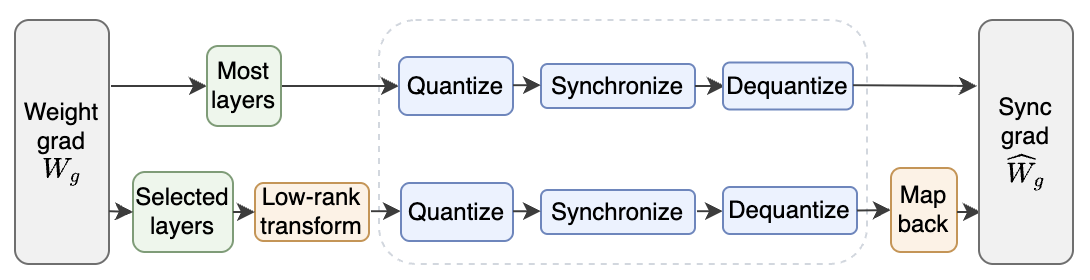}
    \caption{Pretraining-time structure of \name{}. Most layers use the shared communication core directly in Euclidean coordinates. Selected numerically vulnerable layers additionally apply an input-side low-rank transform before entering the core and map the synchronized result back afterward.}
    \label{fig:design_pipeline}
\end{figure}

Figure~\ref{fig:design_pipeline} summarizes the pretraining-time structure of the final \name{} system.
All layers share the same communication core consisting of quantization, communication, and dequantization.
Most layers enter this core directly in Euclidean coordinates, while selected numerically vulnerable layers first apply the input-side low-rank transform and map the synchronized result back afterward.
This organization makes the central systems point explicit: \name{} does not replace the Euclidean baseline globally, but augments it only where geometry is empirically justified.

In this view, the default path for most layers remains identical to the Euclidean baseline.
The geometry-informed branch differs only by adding a transform before the shared FP8 communication core and a map-back step afterward.
Because this branch is activated only on a small subset of layers, the overall system remains close to the baseline in communication structure while correcting the layers most affected by FP8 distortion.

\subsection{Offline Preparation and Pretraining-Time Workflow}
\label{subsec:impl_algorithms}

The final implementation consists of two stages.
First, we run an offline profiling phase to identify the vulnerable layer set.
Second, during pretraining, we activate the geometry-informed branch only on the selected layers, while all other layers continue to follow the Euclidean baseline.
The selected layer set $\mathcal{S}$ is fixed before pretraining, but the input-side geometry factors are refreshed periodically during training.
In our experiments, the factors are updated every 50 training steps.

In the current implementation, the selected layer set $\mathcal{S}$ is fixed before pretraining based on the profiling procedure described above.
This organization keeps the layer-selection logic outside the main training loop and preserves a simple fast path for the majority of layers.
Algorithm~\ref{alg:selective-fp8} summarizes the resulting pretraining-time communication rule.
For the selected \name{} layers, we use local scaling in the transformed coordinates, which gives better empirical fidelity than synchronizing a global scale in our ablation.

\begin{algorithm}[!t]
\caption{Selective \name{} Communication with Input-Side Whitening and Error Feedback}
\label{alg:selective-fp8}
\begin{algorithmic}[1]
\Require Fixed selected layer set $\mathcal S$
\Require For each $l \in \mathcal S$, input-side factorization updated every $K$ steps
\Statex \hspace{\algorithmicindent} $A^{(l)} \approx L_A^{(l)} L_A^{(l)\top}$, with $K=50$ in our experiments
\Require Error buffers $\{R^{(l)}\}_{l \in \mathcal S}$ initialized to zero
\Require World size $N$, FP8 quantizer $\mathcal Q(\cdot;s)$, scaling rule $\mathrm{Scale}(\cdot)$
\For{each training step}
\For{each layer $l$ with weight gradient $W_g^{(l)}$}
\If{$l \notin \mathcal S$}
\State $s^{(l)} \gets \mathrm{Scale}(W_g^{(l)})$
\State $s^{(l)} \gets \textsc{AllReduce}_{\mathrm{MAX}}(s^{(l)})$
\State $Q^{(l)} \gets \mathcal Q(W_g^{(l)};s^{(l)})$
\State $\widehat Q^{(l)} \gets \textsc{AllReduce}_{\mathrm{SUM}}(Q^{(l)})$
\State $\widehat W_g^{(l)} \gets \mathrm{Dequantize}(\widehat Q^{(l)};s^{(l)}) / N$
\Else
\State $U^{(l)} \gets W_g^{(l)} \big(L_A^{(l)}\big)^{-\top}$
\State $\widetilde U^{(l)} \gets U^{(l)} + R^{(l)}$
\State $s_{\mathrm{loc}}^{(l)} \gets \mathrm{Scale}(\widetilde U^{(l)})$
\State $\widetilde Q^{(l)} \gets \mathcal Q(\widetilde U^{(l)};s_{\mathrm{loc}}^{(l)})$
\State $\widetilde U_{\mathrm{deq}}^{(l)} \gets \mathrm{Dequantize}(\widetilde Q^{(l)};s_{\mathrm{loc}}^{(l)})$
\State $R^{(l)} \gets \widetilde U^{(l)} - \widetilde U_{\mathrm{deq}}^{(l)}$
\State $s_{\mathrm{comm}}^{(l)} \gets s_{\mathrm{loc}}^{(l)}$
\If{this is a scale-synchronization step}
\State $s_{\mathrm{comm}}^{(l)} \gets \textsc{AllReduce}_{\mathrm{MAX}}(s_{\mathrm{loc}}^{(l)})$
\EndIf
\State $\widehat Q^{(l)} \gets \textsc{AllReduce}_{\mathrm{SUM}}(\widetilde Q^{(l)})$
\State $\widehat U^{(l)} \gets \mathrm{Dequantize}(\widehat Q^{(l)};s_{\mathrm{comm}}^{(l)}) / N$
\State $\widehat W_g^{(l)} \gets \widehat U^{(l)} L_A^{(l)\top}$
\EndIf
\EndFor
\State Optimizer updates parameters using $\{\widehat W_g^{(l)}\}$
\EndFor
\end{algorithmic}
\end{algorithm}

Two implementation details are worth emphasizing.
First, the geometry-aware branch is activated only on a fixed selected subset of layers, so the majority of the model follows the original Euclidean path without additional geometric processing.
Second, the geometry-aware branch uses error feedback in the transformed coordinates, allowing quantization residuals to be accumulated where the communication actually occurs.
This keeps the implementation aligned with the communication formulation rather than treating error feedback as a separate post hoc correction.

\subsection{Implementation Summary}
\label{subsec:impl_summary}

Our final implementation is obtained by progressively simplifying the full geometry-informed formulation of Section~\ref{sec:design}.
We first remove the output-side transform after observing that the input-side geometry captures nearly all of the practical benefit of the full design.
We then compress the remaining input-side geometry with a rank-32 low-rank approximation.
Finally, we deploy the resulting transform only on a small subset of numerically vulnerable layers.

These simplifications yield a practical hybrid system:
\emph{Euclidean remains the default communication path, and geometry is introduced only where it is empirically justified.}
As a result, the implementation stays close to the baseline in communication structure and cost while recovering most of the measured fidelity benefit of full geometry-aware communication on the layers that need it most.

\FloatBarrier
\section{Experimental Results}
\label{sec:exp}

We evaluate \name{} from four complementary perspectives.
First, we measure the system benefit of low-precision communication as the model size and GPU count increase.
Second, we examine whether \name{} preserves pretraining behavior in terms of validation loss.
Third, we compare downstream task performance to determine whether geometry-aware communication better preserves useful model quality than direct Euclidean communication. Fourth, we examine the memory overhead of \name{} relative to the direct Euclidean baseline.

Our goal is not to claim that \name{} improves every metric uniformly.
Rather, the goal is to test a more specific hypothesis:
\emph{under the same low-precision communication format, changing only the communication coordinates can improve the usefulness of the communicated update.}
This distinction is important in our setting, because optimization loss and downstream task quality do not always rank the methods in the same way.

\subsection{Experimental Setup}
\label{subsec:exp_setup}

We study two LLaMA-style pretraining configurations on OpenWebText.
The first is a model of approximately 300M parameters pretrained with a sequence length of 4096.
The second is a model of approximately 600M parameters pretrained with a sequence length of 2048.
We use different sequence lengths to fit in the GPU memory.

Unless otherwise specified, all end-to-end pretraining experiments use 32 GPUs, global batch size 512, micro batch size 4, Muon optimizer, learning rate $5\times10^{-4}$, cosine decay, and minimum learning rate $5\times10^{-5}$.

All experiments are conducted on the Vista supercomputer at the Texas Advanced Computing Center (TACC).
Vista includes 600 Grace Hopper compute nodes based on the NVIDIA GH200 Grace Hopper Superchip architecture, and each Grace Hopper node provides one NVIDIA GPU with 96\,GB of HBM3 memory together with an NVIDIA Grace CPU interconnected through a tightly integrated CPU--GPU design. 

For end-to-end pretraining, we compare three practical methods:
\begin{itemize}
    \item \textbf{FP32}: full-precision gradient communication.
    \item \textbf{Euclidean baseline}: direct layer-wise FP8 communication in Euclidean parameter coordinates.
    \item \textbf{\name{}}: the final practical geometry-aware design, which applies the geometry-aware branch only where its benefit justifies the cost, while keeping the Euclidean baseline path elsewhere.
\end{itemize}
We choose the stronger baseline in each category based on additional 600M experiments.
BF16 gradient communication matches FP32 on 7 out of 14 downstream tasks, but its absolute task values are slightly lower overall, so we use FP32 as the high-precision reference.
For Euclidean FP8, standard per-block scaling with block size 512 matches FP32 on only 4 out of 14 downstream tasks, which is worse than the layer-wise Euclidean baseline; therefore, we use the stronger layer-wise Euclidean FP8 baseline in the main comparison.

\subsection{Scaling Benefit of Low-Precision Communication}
\label{subsec:exp_scaling}

\begin{figure}[t]
    \centering
    \includegraphics[width=0.96\linewidth]{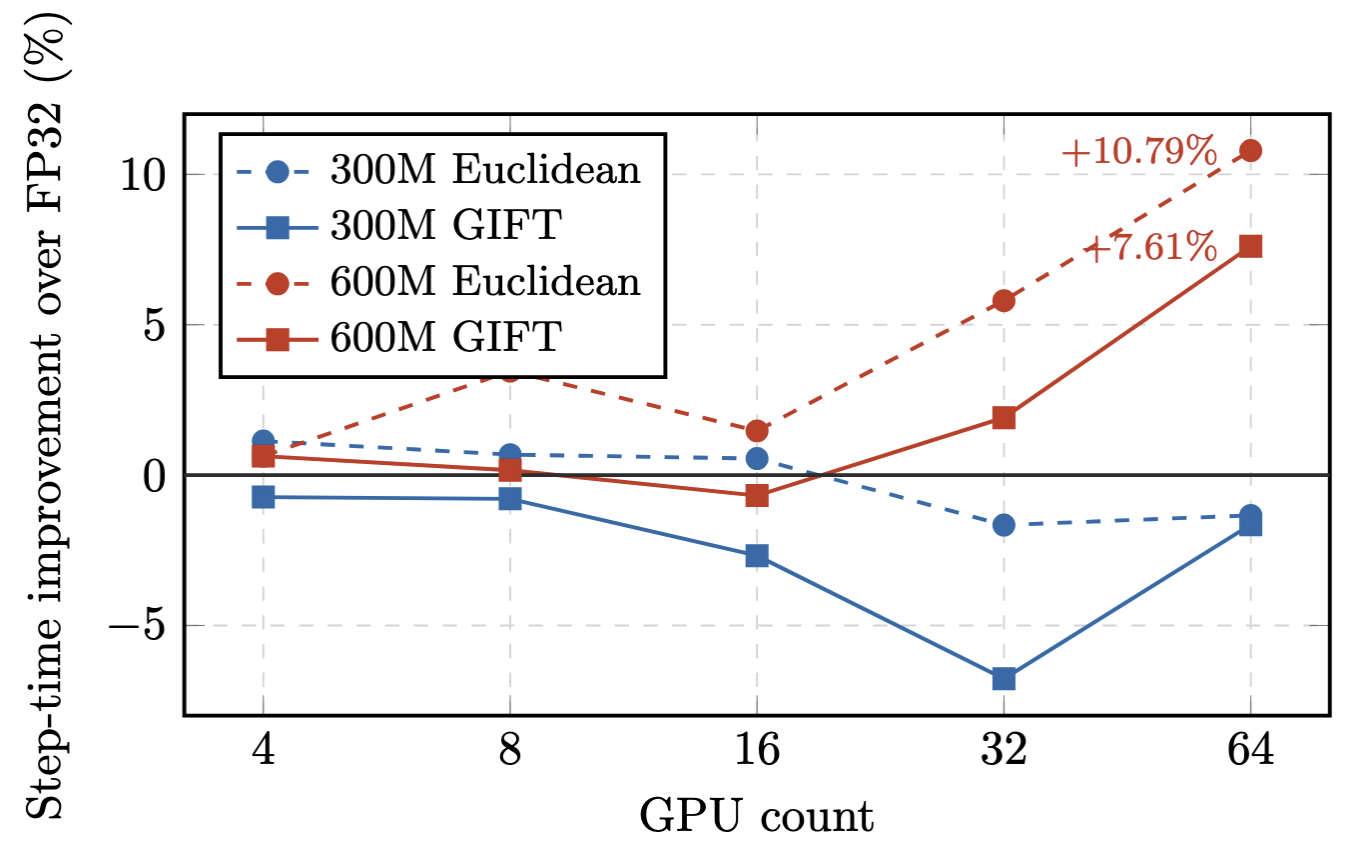}
    \caption{Step-time improvement over FP32 communication as GPU count increases. Positive values indicate faster pretraining than the FP32 baseline; negative values indicate slower pretraining. The relative benefit of low-precision communication becomes more pronounced at larger GPU counts, especially for the 600M model. \name{} is slower than direct Euclidean baseline because of its additional geometry-aware computation, but it retains a substantial fraction of the communication-side speedup over FP32.}
    \label{fig:step_improvement_scale}
\end{figure}

We first examine the system benefit of low-precision communication as scale increases.
Figure~\ref{fig:step_improvement_scale} reports the step-time improvement of Euclidean baseline and \name{} relative to the FP32 baseline as a function of GPU count, for both the 300M and 600M models.
Positive values indicate that the method is faster than FP32, while negative values indicate that it is slower.

We observe three trends.
First, the communication advantage of low-precision communication becomes more pronounced at a larger scale, especially for the 600M model.
Second, the larger model benefits more consistently from reduced communication cost, which is expected because communication occupies a larger fraction of end-to-end step time at a larger scale.
Third, \name{} remains slower than direct Euclidean baseline because it introduces additional geometry-aware computation, but it still retains a substantial fraction of the low-precision systems' benefit relative to FP32.
Most importantly, the gap widens as GPU count increases, which is precisely the regime where communication efficiency matters most.

These results indicate that low-precision communication is more effective for larger models at larger scales, creating room for a geometry-aware method that trades a controlled amount of extra computation for better preservation of model quality.

\subsection{Validation Loss During Pretraining}
\label{subsec:exp_loss}

\begin{figure}[!t]
    \centering
    \includegraphics[width=\linewidth,height=0.15\textheight,keepaspectratio]{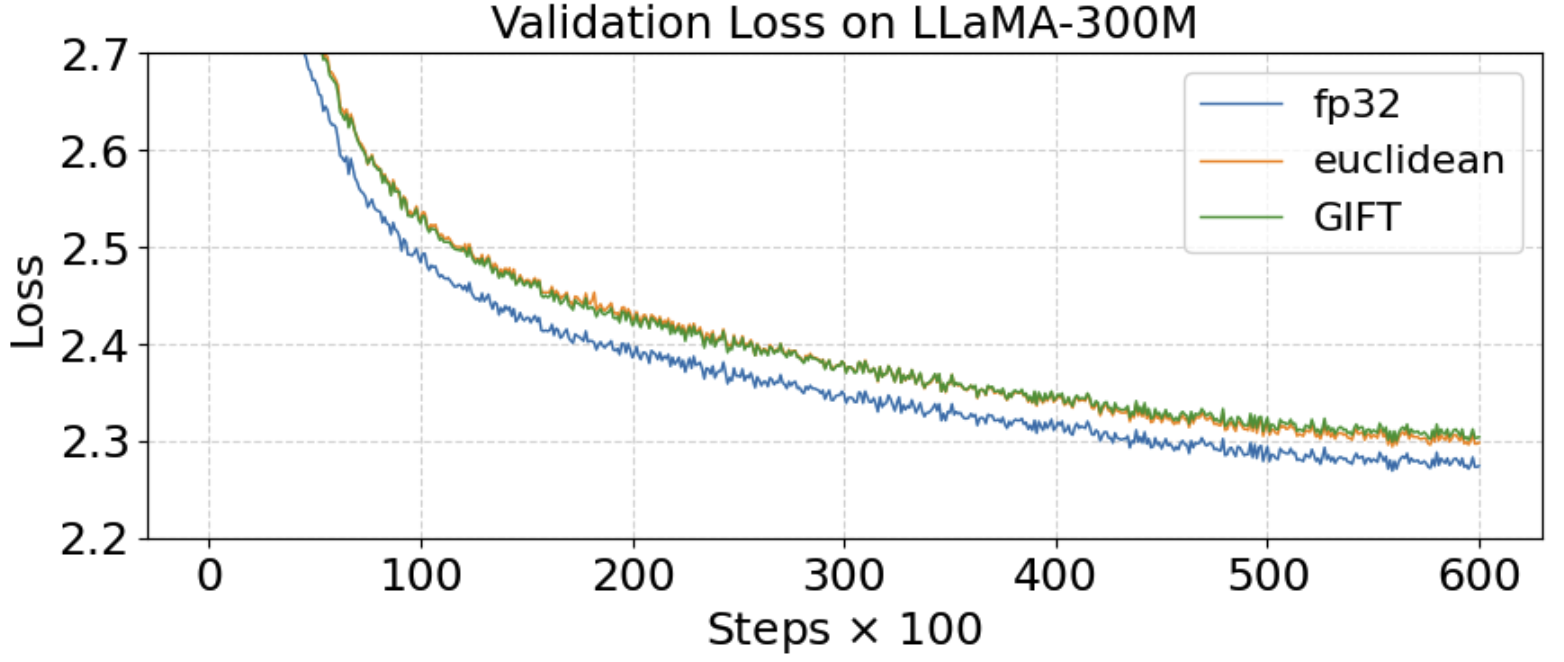}
    \vspace{0.4em}
    
    \includegraphics[width=\linewidth,height=0.15\textheight,keepaspectratio]{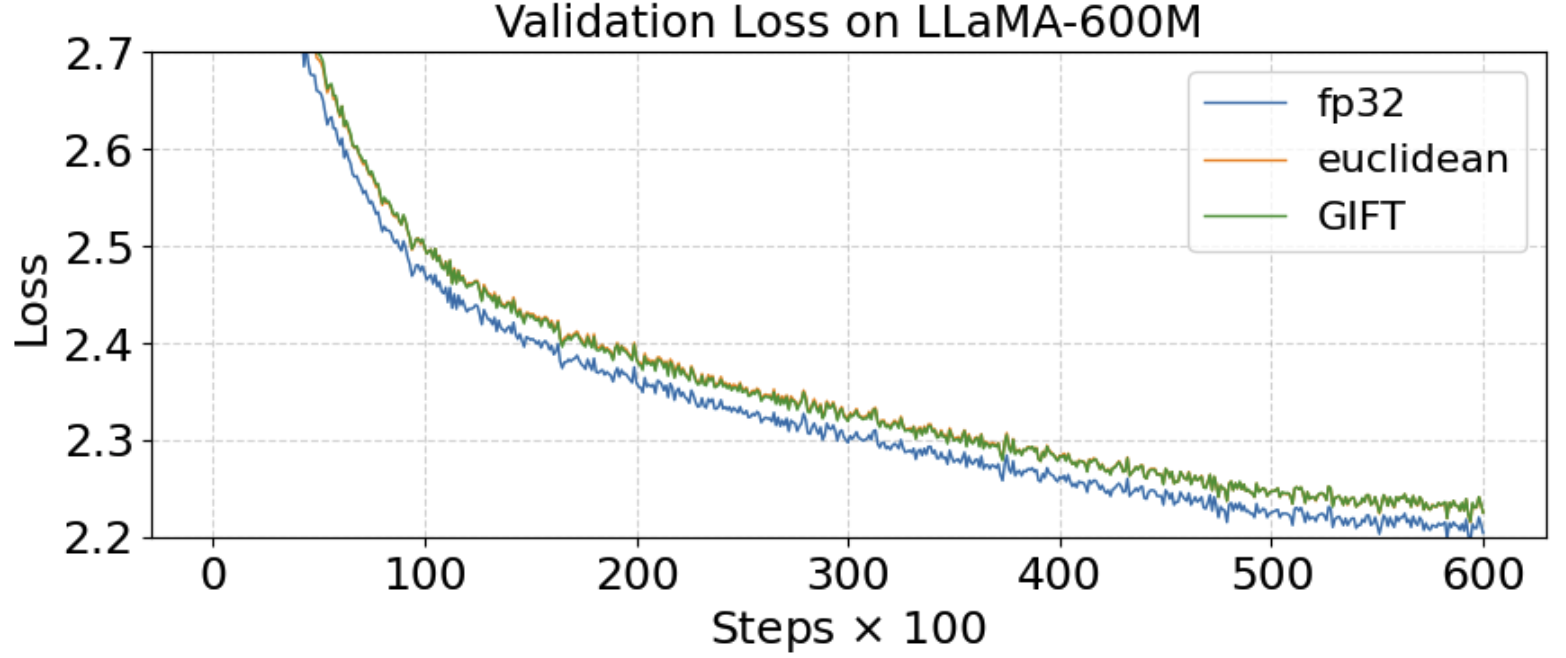}
    \caption{Validation loss during pretraining for the 300M model (top) and the 600M model (bottom), comparing FP32, direct Euclidean baseline communication, and \name{}. In both cases, \name{} and Euclidean baseline exhibit very similar validation-loss curves, with both remaining reasonably close to the FP32 baseline. This also highlights a broader point: validation loss is only a partial proxy for pretraining quality, and differences between communication methods may become more apparent in downstream performance than in pretraining loss alone.}
    \label{fig:val_loss_main}
\end{figure}

We next examine pretraining behavior through validation loss.
Figure~\ref{fig:val_loss_main} compares FP32, Euclidean baseline, and \name{} for both model sizes.

The main observation is that \name{} and Euclidean baseline exhibit very similar validation-loss trajectories across both model sizes.
Both low-precision methods remain reasonably close to the FP32 baseline throughout pretraining, but the gap between \name{} and Euclidean baseline is small in validation loss.

This point is important for interpreting the rest of the results.
Validation loss is a useful optimization-level signal, but our method changes communication fidelity rather than the model architecture or objective.
In such a setting, small differences in validation loss do not necessarily imply degraded downstream model performance.
For that reason, we treat validation loss as an informative but incomplete measure, and use downstream evaluation as the decisive test.

\begin{table*}[t]
\centering
\caption{Downstream task performance of models pretrained with FP32, BF16, Euclidean baseline, per-block Euclidean FP8, full K-FAC, and geometry-informed communication. Higher is better. ``Wins vs.\ FP32'' counts the number of tasks on which the method outperforms the FP32 baseline.}
\label{tab:downstream_main}
\resizebox{\textwidth}{!}{
\begin{tabular}{llccccccccccccccc}
\toprule
Model & Method & BOOLQ & CB-ACC & COPA & MUL-RC & RCD-F1 & RTE & WiC & WSC & LAMB. & RACE & M-QA & PIQA & WINO. & LAMB-STD & Wins vs.\ FP32 \\
\midrule
\multirow{5}{*}{600M}
& FP32      & 0.5908 & 0.2679 & 0.7500 & 0.5532 & 0.8352 & 0.5090 & 0.5000 & 0.4423 & 0.4941 & 0.3120 & 0.2251 & 0.6719 & 0.5612 & 0.3716 & -- \\
& BF16      & 0.6058 & 0.2321 & 0.7100 & 0.5608 & 0.8346 & 0.5307 & 0.4875 & 0.4231 & 0.4865 & 0.3158 & 0.2295 & 0.6806 & 0.5391 & 0.3873 & 7/14 \\
& Per-block Euclidean & 0.5679 & 0.1250 & 0.7100 & 0.5549 & 0.8205 & 0.5487 & 0.5188 & 0.6058 & 0.4795 & 0.3024 & 0.2161 & 0.6665 & 0.5478 & 0.3685 & 4/14 \\
& Euclidean & 0.6119 & 0.4107 & 0.7100 & 0.5025 & 0.8271 & 0.5199 & 0.4969 & 0.5865 & 0.4824 & 0.3100 & 0.2201 & 0.6659 & 0.5320 & 0.3850 & 5/14 \\
& \name{}        & 0.6104 & 0.3036 & 0.7200 & 0.5054 & 0.8197 & 0.5307 & 0.5031 & 0.4712 & 0.4875 & 0.3033 & 0.2157 & 0.6763 & 0.5257 & 0.3722 & 7/14 \\
\midrule
\multirow{4}{*}{300M}
& FP32       & 0.5226 & 0.4107 & 0.6900 & 0.5528 & 0.8020 & 0.5271 & 0.5031 & 0.3846 & 0.4514 & 0.2986 & 0.2261 & 0.6507 & 0.5375 & 0.3223 & -- \\
& Euclidean  & 0.5734 & 0.2857 & 0.6500 & 0.4878 & 0.7954 & 0.5271 & 0.5016 & 0.3654 & 0.4347 & 0.3043 & 0.2224 & 0.6480 & 0.5162 & 0.3334 & 4/14 \\
& Full K-FAC & 0.6064 & 0.4107 & 0.7000 & 0.5140 & 0.7799 & 0.5343 & 0.5063 & 0.3654 & 0.4225 & 0.3081 & 0.2188 & 0.6458 & 0.5335 & 0.3076 & 6/14 \\
& \name{}         & 0.5740 & 0.2321 & 0.7200 & 0.4420 & 0.7881 & 0.5848 & 0.4969 & 0.4038 & 0.4256 & 0.3014 & 0.2302 & 0.6621 & 0.5335 & 0.3089 & 7/14 \\
\bottomrule
\end{tabular}
}
\end{table*}

\subsection{Downstream Task Performance}
\label{subsec:exp_downstream}

We now turn to the main practical question of the paper:
Which method better preserves useful model quality after end-to-end pretraining?

Table~\ref{tab:downstream_main} reports downstream results for all evaluated methods.
Each column corresponds to one downstream task, and the rightmost column summarizes the number of task wins relative to FP32.
We present the results task by task because the main question is not only overall performance, but also how consistently each communication method preserves useful model quality across different downstream evaluations.

The downstream results reveal a clearer distinction than the validation-loss curves.
For the 300M model, Euclidean baseline outperforms FP32 on 4 out of 14 tasks, whereas \name{} outperforms FP32 on 7 out of 14 tasks.
We also evaluate the full K-FAC variant on the 300M model.
Although full K-FAC uses a more complete geometry-aware state, it outperforms FP32 on 6 out of 14 tasks, which is still worse than the final selective \name{} design.
This supports our design choice that applying geometry awareness selectively can preserve more useful downstream behavior than using a heavier full K-FAC-style branch everywhere.

For the 600M model, Euclidean baseline outperforms FP32 on 5 out of 14 tasks, whereas \name{} again outperforms FP32 on 7 out of 14 tasks.
We additionally report the results for the 600M BF16 and per-block Euclidean FP8 ablations.
BF16 matches FP32 on 7 out of 14 tasks, but its absolute task values are slightly lower overall, so FP32 remains the stronger high-precision reference.
Per-block Euclidean FP8 matches FP32 on only 4 out of 14 tasks, which is worse than the layer-wise Euclidean baseline.
Therefore, the main Euclidean comparison uses the stronger layer-wise FP8 baseline rather than the weaker per-block variant.

Thus, while all methods remain relatively close in validation loss, \name{} yields a more favorable downstream profile than direct Euclidean baseline across both model sizes and also improves over the heavier full K-FAC variant in the 300M ablation.

These results should not be interpreted as \name{} matching FP32 on every task.
Rather, they show that \name{} improves the cross-task preservation profile over direct Euclidean FP8 communication.
The purpose of communication compression is not merely to keep the loss curve visually close to FP32, but to preserve the usefulness of the synchronized update.
Our results indicate that geometry-aware communication better protects task-relevant pretraining signal under aggressive low-precision communication.
In other words, changing the communication coordinates improves the downstream preservation profile over the Euclidean baseline, even when the improvement is only weakly reflected in validation loss.

\subsection{Memory Study}
\label{subsec:memory}
We compare the memory footprint of \name{}  with that of the direct Euclidean baseline.
For the 300M model, \name{} increases memory usage by approximately 3.33\%.
For the 600M model, the increase is approximately 8.98\%.
This overhead mainly comes from the stored input-side low-rank factors and transformed-coordinate error-feedback buffers for the selected layers. Since the geometry branch is selective, the overhead remains limited rather than scaling as a full two-sided per-layer K-FAC state.
Overall, these results show that the geometry-aware design introduces only a modest memory overhead for the two models.
\subsection{Discussion}
\label{subsec:exp_discussion}

The experiments support three main conclusions.

First, the system benefit of low-precision communication becomes larger as the scale increases.
This is especially clear for the 600M model at higher GPU counts, where communication occupies a larger fraction of overall step time.

Second, both \name{} and the direct Euclidean baseline remain reasonably close to FP32 in validation loss across both model sizes, while also exhibiting very similar loss trajectories to each other.
This suggests that validation loss is an informative but incomplete proxy for the effect of communication quality on final model performance.

Third, and most importantly, downstream evaluation shows that geometry-aware communication better preserves model quality than direct Euclidean baseline communication.
Across both models, \name{} achieves a more favorable downstream preservation profile than the Euclidean baseline, although it does not improve every individual task.
The 300M full K-FAC ablation and the 600M BF16 and per-block Euclidean ablations further support the choice of the final comparison setting: full K-FAC is not better than the selective design, BF16 is not a stronger reference than FP32, and per-block Euclidean FP8 is weaker than the layer-wise Euclidean baseline.
This means the main benefit of \name{} would be understated if evaluation were restricted to loss curves alone.

Taken together, these results support the central claim of the paper:
Changing communication coordinates can make low-precision gradient communication more faithful to the high-precision reference than direct Euclidean FP8, and with selective deployment, this can be done with a practical systems cost.
Although our experiments focus on FP8, the coordinate-change principle behind \name{} is not specific to FP8, suggesting that the same idea can be applied to future FP4 communication once hardware and software support are mature.
\subsection{Limitations}
\label{subsec:exp_limitations}

Our current evaluation has several limitations.
First, the end-to-end experiments focus on two medium-scale Llama models and a fixed pretraining recipe.
It would be valuable to test larger models, longer training horizons, and broader hardware regimes.
Second, although the downstream results already show a consistent advantage of \name{} over direct Euclidean baseline, the evaluation would be further strengthened by additional seeds and a larger benchmark suite.
Third, this work focuses specifically on communication-only low-precision communication rather than a fully quantized pretraining stack.
Future work can study whether the same geometry-aware principle remains useful when combined with broader low-precision pretraining recipes.

Even with these limitations, the present results already establish the main practical point:
geometry-aware communication improves the quality retained under FP8 gradient communication relative to direct Euclidean FP8, while preserving part of the communication-side efficiency that makes low-precision communication attractive.

\section{Conclusion}
\label{sec:conclusion}

This paper studies low-precision gradient communication for LLM pretraining from a geometry-informed perspective.
Our starting point is the observation that low-precision communication error is not only a numerical-format issue, but also a coordinate-system issue:
Under anisotropic gradient geometry, direct quantization in Euclidean coordinates can introduce substantial distortion into the aggregated update.
Motivated by this insight, we formulate a full geometry-aware communication rule based on a K-FAC-style second-order structure and progressively simplify it into a practical system implementation.
Our results show that most of the measured FP8 round-trip fidelity benefit of the full two-sided formulation can be recovered by a much cheaper hybrid design that retains only the input-side geometry, approximates it with a 32-rank representation, and applies it selectively to the most numerically vulnerable layers.
The resulting \name{} system preserves the Euclidean baseline fast path for most of the model while improving the fidelity–overhead tradeoff of low-precision communication.
\name{} achieves a 7.6\% pretraining time improvement for Llama-600M pretraining across 64 NVIDIA GH200 Superchips, while improving the downstream preservation profile over direct Euclidean FP8 communication.
As communication pressure continues to grow with model size and GPU scale, geometry-aware communication may become increasingly useful for future lower-bit LLM pretraining systems.

We hope this perspective encourages further work on communication-aware geometric methods for efficient distributed LLM pretraining.

\balance

\bibliographystyle{IEEEtran}
\bibliography{refs}

\end{document}